\documentclass[journal=jacsat,manuscript=article]{achemso}
\setkeys{acs}{keywords = true}

\usepackage[version=3]{mhchem} 
\usepackage[dvipsnames]{xcolor}
\usepackage{siunitx}
\usepackage{xcolor,colortbl}
\usepackage{chapterbib}

\definecolor{Gray}{rgb}{0.47, 0.7, 1}
\definecolor{LightGray}{rgb}{0.6, 0.8, 1}

\usepackage{empheq}
\usepackage[most]{tcolorbox}

\newtcbox{\mymath}[1][]{%
    nobeforeafter, math upper, tcbox raise base,
    enhanced, colframe=blue!10!black,
    colback=blue!15, boxrule=1pt,
    #1}

\author{Laura N. Casses}
\affiliation[DTU]
{Department of Photonics Engineering, Technical University of Denmark, 2800 Kongens Lyngby, Denmark}
\alsoaffiliation[CNG]
{Centre for Nanostructured Graphene, Technical University of Denmark, 2800 Kongens Lyngby, Denmark}
\alsoaffiliation[Nanophoton]
{Centre for Nanophotonics, Technical University of Denmark, 2800 Kongens Lyngby, Denmark}

\author{Korbinian J. Kaltenecker}
\altaffiliation{Current address: Attocube Systems AG, Eglfinger Weg 2, 85540 Haar, 
Germany}

\author{Sanshui Xiao}
\affiliation[DTU]
{Department of Photonics Engineering, Technical University of Denmark, 2800 Kongens Lyngby, Denmark}
\alsoaffiliation[CNG]
{Centre for Nanostructured Graphene, Technical University of Denmark, 2800 Kongens Lyngby, Denmark}
\alsoaffiliation[Nanophoton]
{Centre for Nanophotonics, Technical University of Denmark, 2800 Kongens Lyngby, Denmark}

\author{Martijn Wubs}
\affiliation[DTU]
{Department of Photonics Engineering, Technical University of Denmark, 2800 Kongens Lyngby, Denmark}
\alsoaffiliation[CNG]
{Centre for Nanostructured Graphene, Technical University of Denmark, 2800 Kongens Lyngby, Denmark}
\alsoaffiliation[Nanophoton]
{Centre for Nanophotonics, Technical University of Denmark, 2800 Kongens Lyngby, Denmark}

\author{Nicolas Stenger}
\affiliation[DTU]
{Department of Photonics Engineering, Technical University of Denmark, 2800 Kongens Lyngby, Denmark}
\alsoaffiliation[CNG]
{Centre for Nanostructured Graphene, Technical University of Denmark, 2800 Kongens Lyngby, Denmark}
\alsoaffiliation[Nanophoton]
{Centre for Nanophotonics, Technical University of Denmark, 2800 Kongens Lyngby, Denmark}
\email{niste@fotonik.dtu.dk}

\title[An \textsf{achemso} demo]{Quantitative near-field characterization of surface plasmon polaritons on monocrystalline gold platelets}
  
\begin{document}

\begin{abstract}
 The subwavelength confinement of surface plasmon polaritons (SPPs) makes them attractive 
 for various applications such as sensing, light generation and solar energy conversion.
 Near-field microscopy associated with interferometric detection allows to visualize both the amplitude and phase of SPPs. However, their full quantitative characterization in a reflection configuration is challenging due to complex wave patterns arising from the interference between several excitation channels. Here, we present near-field measurements of SPPs on large monocrystalline gold platelets in the visible spectral range. We study systematically the influence of the incident angle of the exciting light on the SPPs launched by an
 atomic force microscope tip. We find that the amplitude and phase signals of these SPPs are best disentangled from other signals at grazing incident angle relative to the edge of the gold platelet. Furthermore, we introduce a simple model to explain the $\pi/2$ phase shift observed  between the SPP amplitude and phase profiles. Using this model, the wavelength and propagation length of the tip-launched plasmons are retrieved by isolating and fitting their signals far from the platelets' edges. Our experimental results are in excellent agreement with theoretical models using gold refractive index values. The presented method to fully characterize the SPP complex wavevector could enable the quantitative analysis of polaritons occurring in different materials at visible wavelengths.


\end{abstract}



\section{Introduction}
Surface plasmon polaritons (SPPs) can confine light in small volumes at the subwavelength scale.\cite{Maier2007plasmonics}
SPPs are surface waves and are being investigated for various applications such as sensing\cite{Roh2011overview}, nanoplasmonic circuits\cite{Huang2010atomically}, light generation \cite{Schmidt2012adiabatic,Kern2015electrically} and photovoltaics \cite{Ferry2010design}. 
To design and fabricate the optimal structures for these applications, it is important to know
the complex wavevector, as its real part is inversely proportional to the plasmonic wavelength and its imaginary part inversely proportional to the propagation length\cite{Maier2007plasmonics}.

Scattering-type scanning near-field optical microscopy (s-SNOM) \cite{Keilmann2004near} allows to explore
surface waves with a spatial resolution down to a few tens of nanometers. In the past years, s-SNOM has been demonstrated to be a powerful
tool to study polaritons of various kinds such as polaritons in two-dimensional (2D) materials \cite{Basov2016polaritons,Low2017polaritons} throughout the optical spectrum\cite{Hu2017imaging,Chen2012optical,Fei2012gate,Dai2014tunable,Woessner2015highly,Bylinkin2021real}, 
in anisotropic materials \cite{Ma2018plane,Taboada2020broad} and SPPs on metals \cite{Andryieuski2014direct,Walla2018anisotropic,Kaltenecker2020mono}. 
While the characterization of polaritons has been done in many works in the mid-infrared range\cite{Chen2012optical,Fei2012gate,Dai2014tunable,Woessner2015highly,Bylinkin2021real}, s-SNOM studies in the visible range are less common. s-SNOM measurements in the visible are indeed more challenging because the smaller extension of the light spot focused onto the atomic force microscope (AFM) tip makes the alignment more difficult and renders less stable scans. The study of surface waves in the visible is nevertheless relevant, as many polaritonic phenomena arise in this range \cite{Basov2021polariton,Low2017polaritons,Basov2016polaritons,Hu2017imaging}. 
In the case of SPPs on gold, the characterization of plasmonic slot waveguides has been achieved with a s-SNOM at telecom wavelengths\cite{Andryieuski2014direct,Pramassing2020interferometric} and the direct characterization of SPPs on a flat surface with an aperture-type SNOM has been achieved at 780 nm\cite{Sierant2021near}. However, these works measured the SPPs in a transmission configuration, where the sample is illuminated from the bottom and the AFM tip is used as a local scatterer to collect the near-field signal.
This configuration is thus limited to substrates that are transparent at the excitation wavelength. 
By contrast, the characterization of SPPs in the reflection configuration offers more flexibility in the choice of substrate. Moreover, the transmission configuration requires the mediation of a well-defined scatterer to excite the SPPs, whereas in reflection the AFM tip itself acts as a scatterer to excite the SPPs.  
However, the characterization in reflection comes with the drawback that many excitation pathways are involved, which makes the near-field signal more difficult to analyze.
One of these excitation paths is mediated by the sharp tip apex that provides the wavevector required to excite SPPs. These SPPs are called the tip-launched SPPs. 
Another excitation path is mediated by the sharp edges of the platelet that can also provide the required momentum for SPPs. SPP excitation by the edges is possible in gold since the SPPs have a weak confinement, in contrast to graphene\cite{Chen2012optical,Fei2012gate,Woessner2015highly}. 
In previous works\cite{Walla2018anisotropic,Kaltenecker2020mono}, several SPP excitation pathways involving the platelet edge and the tip shaft have been identified. The SPPs excited or scattered at the edge are called edge-launched SPPs, and the ones reflected at the tip shaft are called tip-reflected edge-launched SPPs. All these SPPs can be identified through their interference patterns, and we showed in a previous study\cite{Kaltenecker2020mono} that the wavelength of the tip-launched SPPs can be retrieved by isolating a region of the platelet free from edge-launched SPPs. However, the SPP propagation length is still to be quantified to fully characterize SPPs in the visible range and in reflection. 

When analyzing surface waves in the reflection configuration, several studies\cite{Hu2017imaging,Chen2012optical,Dai2014tunable,Woessner2015highly,Ma2018plane,Taboada2020broad,Bylinkin2021real} use the pseudo-heterodyne interferometric method\cite{Ocelic2006pseudoheterodyne} to suppress the far-field background and to obtain both the amplitude and phase of the near-field from their measurements. However, the full characterization of the surface waves is often done using either the real part of the field\cite{Woessner2015highly,Ma2018plane}, where the amplitude and phase information have been merged, or only the near-field amplitude\cite{Hu2017imaging,Taboada2020broad}.
The independent use of the phase information as well as a simple model that could be used for the independent fitting and the comparison of the near-field amplitude and phase signals seem thus to be lacking. 

Here, we perform near-field reflection measurements at visible wavelengths on monocrystalline gold platelets \cite{Wu2015single,Krauss2018,Boroviks2018interference,Boroviks2019use,Boroviks2021anisotropic,Munkhbat2021tunable,Frank2017short}, that have shown to host better plasmonic properties compared to the common polycrystalline gold surfaces\cite{Mejard2017advanced,Hoffmann2016new}. 
We isolate the clearest tip-launched SPPs when illuminating our sample with a laser beam impinging with a low azimuthal angle relative to the edge of the gold platelet. With the signal of the tip-launched SPPs, we develop a simple model for the amplitude and phase information, allowing us to determine with high confidence the wavelength as well as the propagation length of SPPs propagating at the surface of a monocrystalline gold platelet. To the best of our knowledge, this is the first time that the propagation length has been determined with a s-SNOM in reflection and in the visible. Furthermore, this model explains clearly the $\pi/2$ phase shift between the amplitude and phase signals observed in a previous study\cite{Li2014origin}. 
This full characterisation of the complex SPP wavevector paves the way towards a broader understanding of the polaritonic phenomena in diverse materials\cite{Basov2021polariton,Low2017polaritons,Basov2016polaritons,Hu2017imaging}. Moreover, the proposed model helps to fundamentally understand the behaviour of s-SNOM amplitude and phase when studying surface waves in the reflection configuration.


\section{Results and discussion}
\begin{figure} [H]
  \centering
  \includegraphics[width=\textwidth]{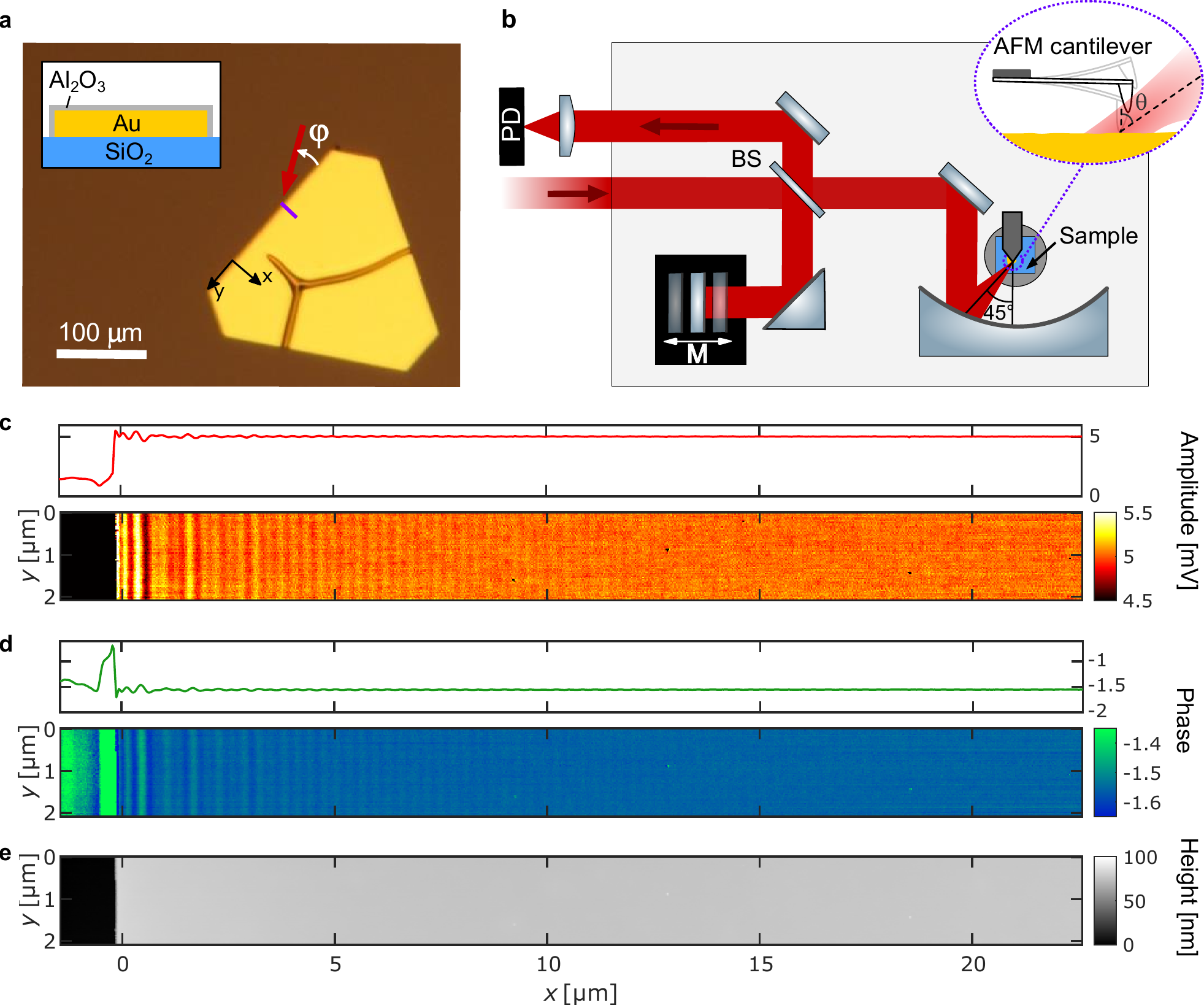}
  \caption{Experimental setup and measurements. (a) Optical image of the sample under study. The incident light is represented by the red arrow and the azimuthal incident angle is represented by $\varphi$. The small violet rectangle close to the tip of the arrow indicates the scanned region. The inset displays a sketch of the sample and substrate cross-section. (b) Sketch of the s-SNOM setup. (c) Near-field amplitude and (d) near-field phase of the scanned region. The respective profiles are shown above each map and the origin of the $x$-axis is chosen to be at the platelet edge. (e) Topography of the scanned region.}
  \label{fig:1}
\end{figure}

\textbf{Measurement of SPPs on gold platelets.}
An optical image of the sample measured in this study is shown in Figure \ref{fig:1}a. A sketch of its cross section is given in the inset. It consists of a monocrystalline gold platelet, synthesized by a wet-chemical synthesis process\cite{Krauss2018} and deposited on a 300 nm thermally grown silicon oxide layer. The platelet under study is about \SI{200}{\micro\meter} large. The propagation length of SPPs on gold is typically an order of magnitude smaller\cite{McPeak2015} than this size. Thus, interference between SPPs reflected from different edges can be avoided simply by scanning a region far from the platelet corners. A 2-nm aluminum oxide (Al$_2$O$_3$) layer has been deposited on the platelet by atomic layer deposition to protect the gold from impurities. The azimutal angle $\varphi$ between the gold platelet's edge and the incident light is indicated in the figure. 

The s-SNOM setup used to characterize the SPPs is represented in Figure \ref{fig:1}b. The light from a helium-neon laser at a wavelength of 633 nm is separated by a beam-splitter (BS) in two paths
of equal intensity. One is focused on an AFM tip oscillating above the sample. The other is directed towards a reference mirror oscillating at a frequency $M = 300$~Hz, necessary for the pseudo-heterodyne interferometric detection \cite{Ocelic2006pseudoheterodyne}. The field scattered at the tip then interferes with the reference field and is detected by a photodiode (PD). Demodulating the total signal enables to retrieve the amplitude and phase of the near-field, as well as to strongly suppress the interference with the far-field background\cite{Ocelic2006pseudoheterodyne}.

The topography, amplitude and phase maps resulting from one of our measurements are displayed in Figures \ref{fig:1}c-e. The topography presented in Figure \ref{fig:1}e gives a sample thickness of about 90 nm. The platelets are thus thick enough to allow us to neglect any possible hybridization between modes propagating at the air/gold and silica/gold interfaces\cite{Grossmann2021micro,Kaltenecker2020mono}. The sample root-mean-square (RMS) roughness extracted from this Figure \ref{fig:1}e is about 200 pm, attesting for
a very high surface quality.
Note that the platelet edge - placed at the $x$-axis origin - is sharp and without defect: this is important to avoid additional perturbations of the propagating wavefront pattern. 
The near-field optical amplitude and phase are shown respectively in Figures \ref{fig:1}c and \ref{fig:1}d. The scales have been adjusted to visualize the SPP interference patterns better. Line profiles are extracted from these maps by averaging the signal over 200 pixels in the $y$-direction. The respective profiles are displayed above the maps. On the gold surface, SPP interference patterns superimposed on a constant field offset are visible up to about 15 \textmu m, both in the near-field amplitude and the near-field phase maps. 

\begin{figure} [H]
  \centering
  \includegraphics[width=\textwidth]{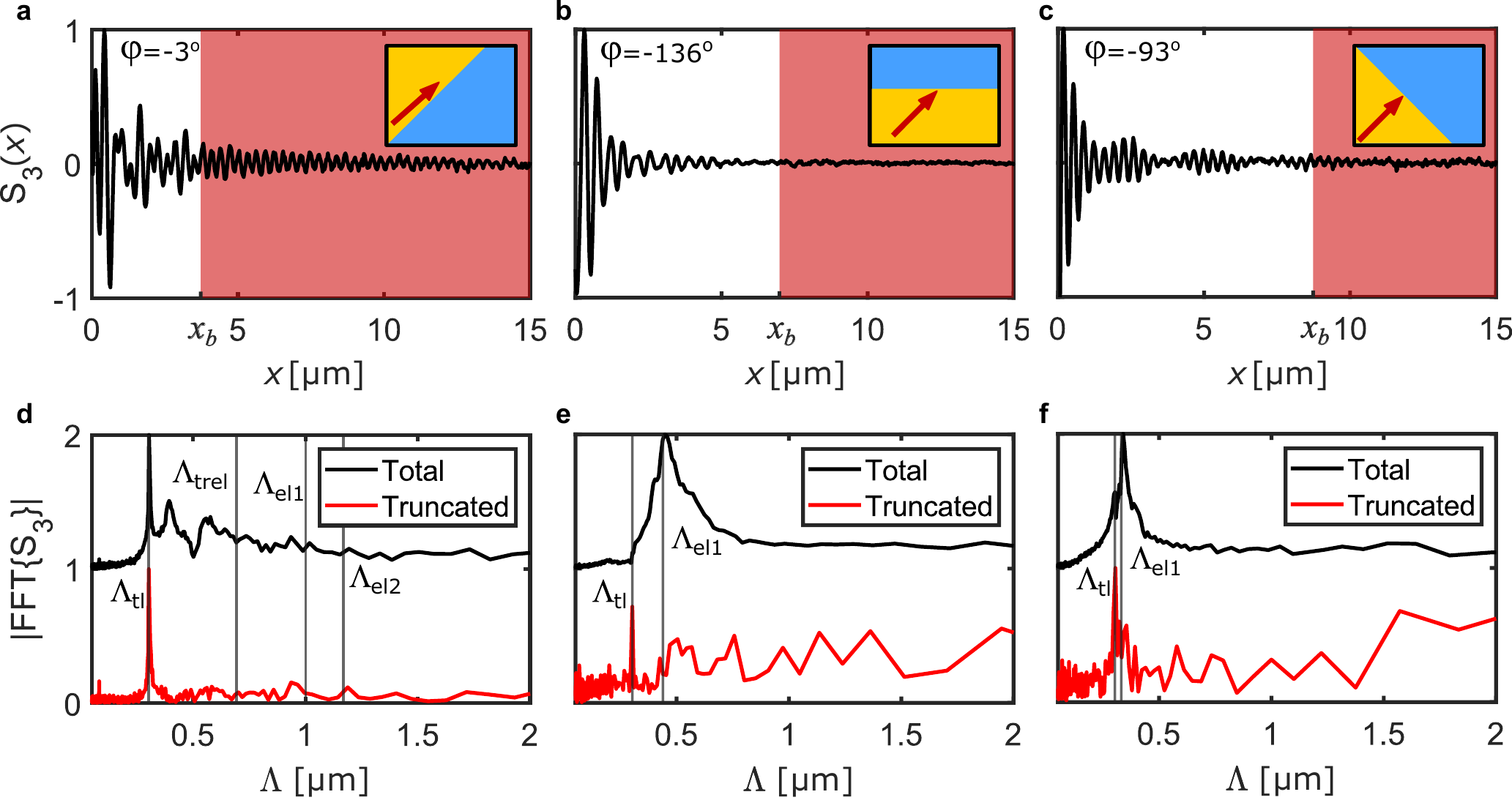}
  \caption{Angle dependency of the different SPPs. (a)-(c) Near-field amplitude profiles of the scanned region on the gold platelets at grazing angle ($\varphi = -3^{\circ}$), at $\varphi = -136^{\circ}$ and at
  $\varphi =-93^{\circ}$, respectively. The red areas starting at the value $x_b$ indicate the portion of the curve where the tip-launched plasmons dominate the signal. The insets depict a top view of the sample edge for each measurement. The red arrow represents the incident light. (d)-(f) represent the  fast Fourier transforms (FFT) of (a)-(c), respectively, as a function of the fringe spacing $\Lambda = 2\pi/K$. The red curve is the FFT of the data in the red area of (a)-(c), demonstrating a clear tip-launched plasmon peak at grazing angle.}
  \label{fig:2} 
\end{figure}

\textbf{Influence of the light incidence} 
Figures \ref{fig:2}a-c present the profiles extracted from three measurements where the azimutal angle $\varphi$ has been varied, with $\varphi = -3^{\circ}$, $-136^{\circ}$ and $-93^{\circ}$, respectively. For clarity, the profiles are normalized and shown on a restricted interval of 15 \textmu m. 
The measurements of three other angles can be found in the section S1 of the Supplementary Information. The three figures show a region closer to the edge with a complex interference pattern, and another region - highlighted by the red background - with more regular damped sinusoidal oscillations. The first region is characterized by the interference between the tip-launched SPPs, the edge-launched SPPs and the tip-reflected edge-launched SPPs\cite{Kaltenecker2020mono}. Because the edge-launched and tip-reflected edge-launched SPPs are present only when the incident light illuminates the edge, the length of the edge region (white background) is taken as always larger than the width of 
the diffraction-limited spot of the incoming light projected onto the sample at a polar angle of $\theta = 60^{\circ}$. 
The Fourier transforms of the amplitude profiles are displayed as the black curves in Figures \ref{fig:2}d-f. To make the small fringe spacing more apparent, the resulting spectra are plotted as a function of the fringe spacing $\Lambda=2\pi/K$ instead of the spatial frequency components $K$.
By considering only the interference of each type of SPP with a strong constant field and neglecting higher-order interference effects such as the interference between the different SPP paths\cite{Kaltenecker2020mono}, several peaks can be recognized. The peaks at the shortest wavelength are the peaks of interest corresponding to the tip-launched SPPs at a wavelength $\Lambda_{tl}$, given by\cite{Kaltenecker2020mono,Walla2018anisotropic,Woessner2015highly}
\begin{equation} \label{eTL}
    \Lambda_\text{tl}  = \frac{\lambda_\text{SPP}}{2} = \frac{\lambda_0}{2 \Re(\Tilde{n})} = \frac{2\pi}{K_\text{tl}}
\end{equation}
where $\Tilde{n}$ is the effective complex refractive index of SPPs at the air/Al$_2$O$_3$/gold interface,  $\Re(\Tilde{n})$ is its real part and $\lambda_0$ is the wavelength of the incident light in vacuum. Since it is directly launched by the tip, $\Lambda_{tl}$ is independent of the incident angle.
Other peaks can be found at the wavelengths of the edge-launched SPPs, $\Lambda_{el1/el2}$ and the wavelength of the tip-reflected edge-launched SPPs, $\Lambda_{trel}$.
Using the phase-matching condition along the platelet edge and approximating the incident wave as a plane wave, these edge-launched SPP wavelengths have been found to be expressed as\cite{Walla2018anisotropic,Kaltenecker2020mono} 
\begin{equation} \label{eEL}
    \Lambda_{el1/el2}(\theta, \varphi) = \frac{\lambda_0}{-\sin(\theta)\sin(\pm\varphi)+\sqrt{\sin^2(\theta)\sin^2(\pm\varphi)-\sin^2(\theta)+\Re(\Tilde{n})^2}},
\end{equation}
where the $+$ sign is used for $\Lambda_{el1}$ and the $-$ sign for $\Lambda_{el2}$. The tip-reflected edge-launched SPP wavelength is expressed as $\Lambda_{trel} = \Lambda_{el1} (\theta',\varphi)$, where $\theta'$ depends on the reflection angle of the incident light at the tip. More information about the different SPP wavelengths and interference paths can be found in our previous work\cite{Kaltenecker2020mono}. To calculate the value of these wavelengths for the different incident angles, we use as reference values the tabulated frequency-dependent refractive index of gold from McPeak \textit{et al.}\cite{McPeak2015} of $\varepsilon_g (\text{633 nm}) = -13.02 + 1.033 i$ and the one of Al$_2$O$_3$ from M. Tulio Aguilar-Gama \textit{et al.}\cite{Aguilar2015structure} of $\varepsilon_{Al_2O_3} (\text{633 nm}) = 2.62$. The value of the gold refractive index has been chosen over other tabulated values\cite{Johnson1972,Olmon2012} as it was giving the longest SPP propagation length at this wavelength. 
Solving the dispersion relation for the multilayer system\cite{Maier2007plasmonics} composed of air, Al$_2$O$_3$ and gold gives the complex SPP wavevector $\beta_{SPP} = (10.37 + 0.03662 i)$ \textmu m$^{-1}$, thus $\Re(\Tilde{n})= \Re(\beta_{SPP}/k_0)=1.045$. Using this last value and Equations \ref{eTL} and \ref{eEL}, the different SPP wavelengths can be determined. These wavelengths are displayed as vertical lines in Figures \ref{fig:2}d-f. In Figure \ref{fig:2}d - corresponding to the grazing incident angle - the tip-launched peak at $\Lambda_{tl}^\text{th}=0.303$ \textmu m is very prominent, while the peaks at about $\Lambda_{el1}(60^\circ,-3^\circ) = 1.0$ \textmu m, $\Lambda_{el2}(60^\circ,-3^\circ) = 1.17$ \textmu m and $\Lambda_{trel}(35^\circ,-3^\circ) = 0.71$ \textmu m can barely be recognized.
However, for incident angles of $\varphi=-136^\circ$ in Figure \ref{fig:2}e and $\varphi=-93^\circ$ in Figure \ref{fig:2}f, peaks at $\Lambda_{el1}(60^\circ,-136^\circ) = 0.44$ \textmu m and $\Lambda_{el1}(60^\circ,-93^\circ) = 0.33$ \textmu m, as well as a smaller tip-launched peak at $\Lambda_{tl}^\text{th}$, can be identified in the corresponding Fourier transforms. 
The large linewidth of the edge-launched contribution can be explained by the fact that Equation \ref{eEL} is valid for an illumination by a plane wave with concomitant spatially constant intensity. In our case the beam is focused on the tip with a numerical aperture of 0.37, thereby exciting SPPs propagating with different angles with respect to the normal of the edge of the platelet. Furthermore, the intensity of the excitation beam at the edge changes while scanning due to the Gaussian profile of the laser beam.

The Fourier transform of the truncated profiles located in the red-shaded areas in Figures \ref{fig:2}a-c are represented by the red curves in Figures \ref{fig:2}d-e. As has been highlighted recently\cite{Kaltenecker2020mono}, selecting the data far from the edges has the effect of filtering out the contribution from the edge-launched SPPs and isolating the tip-launched SPPs, thus leaving a clearer tip-launched peak. However, the significance of this filtering effect is angle dependent. 
For the angles $\varphi=-136^\circ$ and $\varphi=-93^\circ$, truncating has indeed the effect of highly suppressing the peak at $\Lambda_{el1}$ in Figures \ref{fig:2}e and \ref{fig:2}f. However, compared to the truncated profile at grazing azimutal angle, only a very small and noisy tip-launched peak remains.
The RMS of the noise amplitude for $\Lambda$ between 0.1 and 0.2 \textmu m in the truncated profile of Figure \ref{fig:2}d is indeed about 20 times smaller than the tip-launched peak, while only about 5 to 7 times smaller in Figures \ref{fig:2}e and \ref{fig:2}f. Furthermore, in Figure \ref{fig:2}f, the peak at  $\Lambda_{el1}$ is present close to the tip-launched peak. These two effects hinder the proper fitting of the tip-launched peak, and thus the characterization of the tip-launched plasmons at these incident angles. 

The comparison between the different incident angles highlights that the tip-launched SPPs could be best retrieved using the measurement at grazing angle. This angle dependency could be explained by an anisotropic excitation of the SPPs by the tip. Previous results\cite{Walla2018anisotropic} have indeed reported an angle-dependent generation of the tip-launched SPPs, due to the non-circular symmetry of the commercial tip used in their s-SNOM. In the aforementioned study, the angle for which the tip-launched plasmons could be seen best has been identified as the perpendicular incidence, i.e. $\varphi=90^\circ$. The ideal angle is different in our case, as our tip has a different shape than the one used in Ref. \citenum{Walla2018anisotropic}. Therefore, we assign the difference in the ideal angle to the difference in tip shape. 

As a clear tip-launched plasmon peak is needed for the full characterization of the complex wavevector, in the following we will focus our analysis of the SPP properties on the measurement at grazing angle, in the restricted region where only the tip-launched SPPs are present. 

\begin{figure} [H]
  \centering
  \includegraphics[width=\textwidth]{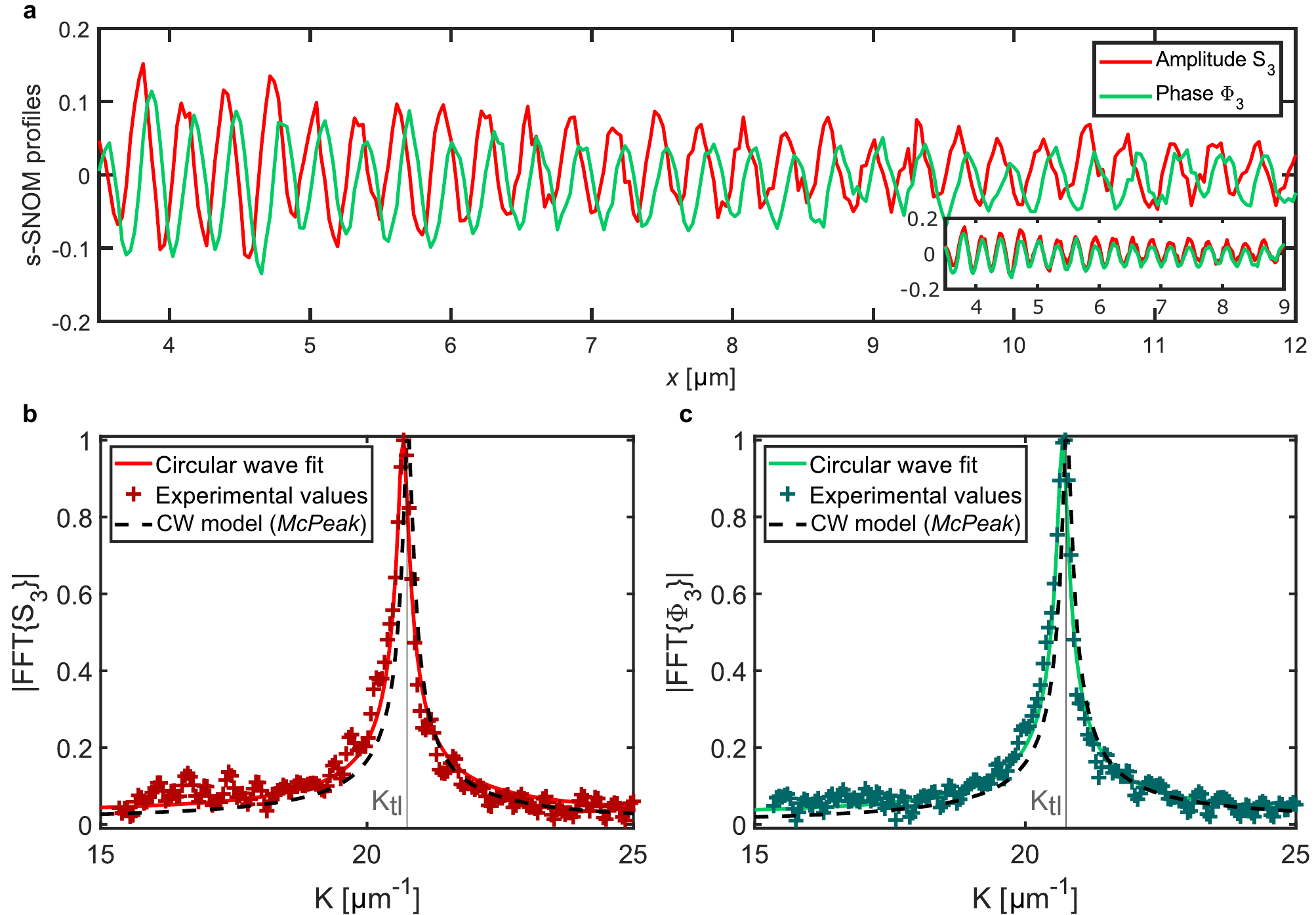}
  \caption{Amplitude and phase at grazing azimutal angle. (a) Comparison of the amplitude and phase profiles at grazing angle. The inset shows the same profiles with the near-field  phase shifted by $-\pi/2$. (b) Fit of the tip-launched SPPs' amplitude peak visible in Fig. \ref{fig:2}(b). The CW model curve refers to the FFT of a circular wave (here Equation \ref{eRA}) where the refractive index from McPeak \textit{et al.}\cite{McPeak2015} has been used. (c) Fit of the corresponding phase data. }
  \label{fig:3}
\end{figure}

\textbf{Model for the detected SPPs.} 
Figure \ref{fig:3}a displays the near-field amplitude and phase profiles of the s-SNOM measurement at grazing azimutal angle $\varphi=-3^\circ$, in the region where only the tip-launched SPPs are present. Both profiles show similar sinusoidal oscillations, with an apparent relative shift between the near-field phase and amplitude by a distance of about $\Lambda_{tl}/4$, corresponding to a phase of $\pi/2$. Besides having been observed in a previous study\cite{Li2014origin}, this $\pi/2$ shift is also confirmed by the good overlap between the two curves when shifting the near-field phase curve by $-\pi/2$, as presented in the inset. 

To explain this phase difference, we introduce a model that considers the interference between the tip-launched SPPs, described as a circular wave, $E_{tl}$, and a constant field $E_g$. 
Because the tip-launched SPPs travel two times the tip-edge distance at each scanning position, the corresponding plasmonic field is described by an apparent wavevector $K_\text{tl} = 2\Re(\beta_{SPP})$ which is twice larger than the actual wavevector, $\Re(\beta_{SPP})$. This also means that we observe an apparent propagation length of the field $L_\text{p}^{\text{tl}} = 1/(2\Im(\beta_{SPP}))$ that is two times smaller than the actual one. 
$E_g=|E_g|e^{i\phi_g}$, with $|E_g|$ its constant amplitude and $\phi_g$ its constant phase, corresponds to the constant offsets seen on the gold surface in the experimental profiles in Figures \ref{fig:1}c and \ref{fig:1}d. Our model does not assume anything about the nature of this constant field. However, as the pseudo-heterodyne detection removes the far-field background due to the incident light\cite{Ocelic2006pseudoheterodyne,Atkin2012nano}, $E_g$ is understood as the near-field coming from the portion of the light back-scattered directly from the sample surface towards the AFM tip\cite{Woessner2015highly}. This effect can be considered as constant because our gold platelets are extremely flat.
With these two contributions, the total near field is $E_{NF} = E_{tl}+E_g$. From the experimental profiles, we observe that typically $|E_g|$ is indeed constant and much larger than the SPP oscillation amplitude $A$ (see the amplitude profile in Figure \ref{fig:1}c). Further analysis 
gives values of $E_g$ that are typically 40 times larger than A. 
In addition, we only consider a range of positions from about 3 \textmu m - to avoid the contributions from the edge SPPs - to about 30 \textmu m - where the oscillations cannot be differentiated from the noise anymore. Using a Taylor expansion in $A/(\sqrt{x}|E_g|)$ of the total amplitude expression and keeping the first-order terms of this Taylor expansion, one can find that (see Supplementary Information Section S2.2)
\begin{equation} \label{eRA}
    |E_{NF}| \approx  |E_g| + \frac{A}{\sqrt{x}} e^{-x/L_\text{p}^{\text{tl}}} \cos{(K_\text{tl}x-\phi)},
\end{equation} 
where $\phi$ is a constant phase consisting of the sum of a plasmonic phase constant $\phi_p$ due to the reflection at the edge and the gold phase constant $\phi_g$.
Using the same principle, the total phase of the near-field can obtained as follows (see Supplementary Information Section S2.3)
\begin{equation} \label{eRP}
    \Phi_{NF} \approx  \frac{A}{|E_g|\sqrt{x}} e^{-x/L_\text{p}^{\text{tl}}} \sin{(K_\text{tl}x-\phi)} + B,
\end{equation} 
where $B$ is a constant depending on the value of $\phi_g$ and on the average optical path difference between the near-field and the reference field\cite{Ocelic2006pseudoheterodyne,Atkin2012nano}. The derivation of these expressions and of the $\pi/2$ relative phase shift can be found in the section S2 of the Supplementary Information. Since the resulting expressions feature a cosine dependence of the total amplitude and a sine dependence of the total phase, the $\pi/2$ shift can be seen as a direct consequence of the interference between the strong constant field and the SPPs.

In addition to highlight in a unambiguous way the $\pi/2$ shift between the near-field amplitude and phase, these expressions have the advantage of being simple enough to be used as fitting functions, meaning that the amplitude and the phase signals can be fitted separately, both in real space and in k-space. While fitting the near-field amplitude or real part of the near-field is fairly common in s-SNOM studies of polaritons\cite{Hu2017imaging,Woessner2015highly,Taboada2020broad}, no independent fitting of the near-field phase has been done so far. As the information extracted from the near-field phase is essentially the same as the near-field amplitude, the two fits can be compared to verify the accuracy of the corresponding parameter values. Note that fitting in k-space has the advantages of displaying the extracted information in a concise manner through a single resonance, and of filtering out the noise at other frequencies. 

Defining $S_3 = |E_{NF}| - |E_g|$ and $\Phi_3 = \Phi_{NF} - B$, the discrete Fourier transform of the near-field data can be fitted. Because the tip-launched SPPs are circular waves and we are using the fast Fourier transform (FFT) algorithm on the measurement profiles, the peak is not expected to be a Lorentzian. An analytical expression of the Fourier transform of a circular wave can be found and has been used to find the SPP wavelength on gold in a previous study\cite{Kaltenecker2021near}. 
However, this analytical expression is not valid for a circular wave profile truncated at the starting value $x_b$ (see Supplementary Information Section S3). For this reason, we chose to fit numerically the signals in k-space. 
The Fourier transform of the experimental data and the circular wave fit are plotted in Figures \ref{fig:3}b and \ref{fig:3}c for the amplitude and phase information, respectively. The fit of the near-field amplitude and phase for the starting value of the truncated segment $x_b = 4.02$ \textmu m, and its comparison with the fit in real space, are summarized in Table \ref{tab:1}. The uncertainties when fitting in k-space are slightly larger, but still comparable to the uncertainties in real space.

\begin{table}[H]
    \centering
    \resizebox{1\textwidth}{!}{
    \begin{tabular}{l ccccc}
        \hline
        \rowcolor{Gray}
        & \multicolumn{2}{c}{\textbf{Amplitude profile}} & \multicolumn{2}{c}{\textbf{Phase profile}} & \textbf{McPeak \textit{et al.}}\\
        \rowcolor{LightGray}
        & k-space & Real space  & k-space  & Real space &   \\
        $K_\text{tl}$ (\textmu m$^{-1}$) & $20.658 \pm 0.009$ & $20.662 \pm 0.003$ & $20.674 \pm 0.007$ & $20.674 \pm 0.009$ & 20.75 \\
        $L_\text{p}^\text{tl}$ (\textmu m) & $12 \pm 1$ & $12.5 \pm 0.4$ & $12 \pm 1$ & $11.9 \pm 0.3$ & 13.65 \\
    \hline
    \end{tabular}
    }
    \caption{Comparison of values for the wavevector $K_\text{tl}=2\pi/\Lambda_\text{tl}$ and propagation length $L_\text{p}^\text{tl}$, as obtained in real space and in Fourier space, for one starting value $x_b$.}
    \label{tab:1}
\end{table}

These values 
could vary with the choice of $x_b$. This starting value has to be sufficiently large to avoid the signal from the edge-launched and tip-reflected edge-launched SPPs, but also sufficiently small to cover as much of the tip-launched SPP signal as possible. Once these two criteria are fulfilled, there still can be some slight changes in the extracted values due to noise that depends on the choice of $x_b$. After studying systematically the evolution of the fitting parameters with $x_b$, we find that our fitting procedure is stable with regard to the choice of $x_b$ (see Supplementary Information Section S4). 

We compare the values of our fitting parameters with theoretical calculations taking into account values of the refractive index for gold and Al$_2$O$_3$ found in the literature\cite{McPeak2015,Aguilar2015structure} (see Table \ref{tab:1}). The FFT of circular waves obtained with these reference values - integrated over the same range as curves of the circular wave fit - are plotted in dashed lines in Figures \ref{fig:3}b and \ref{fig:3}c. While being close, the experimental values are slightly different from the calculated ones, up to about 15$\%$ lower in the case of the propagation length. These discrepancies can be explained by a slightly different experimental refractive index than the one from the literature. Indeed, different studies characterizing gold\cite{McPeak2015,Johnson1972,Olmon2012} have provided slightly different refractive indices. In addition, the refractive index of Al$_2$O$_3$ taken from the literature corresponds to a layer of 50 nm of Al$_2$O$_3$, while we have only 2 nm. This could also induce a difference in the refractive index of our sample. 

We have thus shown that with our method it is possible to extract both the real and imaginary parts of the complex wavevector of our SPPs simultaneously, in excellent agreement with our theoretical model and values found in the literature.

\subsection{Conclusion}

In this work, we have identified the grazing azimutal angle as the best configuration to isolate the tip-launched SPPs. At this particular angle, the near-field amplitude and, for the first time, phase could be fitted. To obtain the expressions for the fitting functions, we have developed a model for the interference between these tip-launched SPPs and a field that we approximated to be a constant. The equations derived for the fitting procedure explicitly justify the damped sinusoidal behaviour of the near-field amplitude and phase, and the $\pi/2$ shift between them. It should be noted that the structures of Equations \ref{eRA} and \ref{eRP} do not depend on the material and are valid as long as the amplitude of the plasmonic oscillations is much smaller than the amplitude of the constant field. Our analysis could thus be applied to a whole range of other materials hosting different types of polaritons. 

With our method, we obtain, for the first time to our knowledge using a s-SNOM in reflection configuration, values of both the gold SPPs wavelength, $\Lambda_\text{tl} = 304,0$ nm, and propagation length, $L_\text{p}^\text{tl} = 12$ \textmu m, corresponding to about 40 coherent oscillations. These values are consistent between the near-field amplitude and phase, do not sensitively depend on the choice of interval used for the fit and are in excellent agreement with previous values found in the literature. Furthermore, our measurements confirm the high surface quality of the monocrystalline gold platelets. With the Al$_2$O$_3$ protection layer, we reproduced our results after eight months (see Supplementary Information Section S4), which shows the stability of our sample. This full characterisation of the SPPs provides a better quantitative understanding of their interactions with other materials. Quantitative study of polaritons in reflection and in the visible is thus facilitated. With the possibility to measure the complex-valued wavevector of polaritons, this work opens the door to the exploration of quantum related phenomena in polariton physics \cite{Boroviks2021extremely,Basov2021polariton,Tame2013quantum} at the nanometer scale, at visible frequencies and irrespective of the type of substrate.

\subsection{Materials and methods}

\hspace*{5mm} \textbf{Mono-crystalline gold platelets.} The gold platelet samples deposited on silicon dioxide are purchased from Nanostruct GmbH. They are fabricated by wet-chemical synthesis, enabling a high aspect ratio of the crystals. The typical RMS flatness obtained with this process is about 200 pm. The platelets are deposited on a silicon dioxide chip and covered by a 2-nm Al$_2$O$_3$ layer by atomic layer deposition.

\textbf{s-SNOM setup.} We use a commercial s-SNOM in reflection mode (neaSNOM, Attocube Systems AG), equipped with a p-polarized 
stabilized HeNe laser (HRS015, Thorlabs GmbH) and a 10-MHz adjustable photodiode (2051, Newport Corp.). The laser is focused by a parabolic mirror on a platinum-coated AFM tip (Arrow-NCPt, NanoAndMore GmbH) with a nominal apex radius of 25 nm. The tip is oscillating above the sample at a frequency of $f_0 \sim 300$ kHz and with an oscillation amplitude of about 60 nm. The signal recorded by the photodiode is demodulated at the first and second sidebands of $3 f_0$\cite{Ocelic2006pseudoheterodyne}. 

\textbf{Fitting method for the k-space fitting:}
The same fitting procedure is applied for the near-field phase and amplitude profiles. First, the profiles truncated from the value $x_b$ to a value of about 30 \textmu m are fitted in real space, using the expressions in Equation \ref{eRA} for the amplitude and Equation \ref{eRP} for the phase, respectively. The constant values in Equations \ref{eRA} and \ref{eRP} are then subtracted. The profile is then zero padded, meaning that a vector of zeros is added at the end of the profile vector. The zero padding allows to obtain an accurate estimate of the peak heights\cite{MatlabZeroP}, and is allowed by the fact that the signal at the end of the amplitude and phase profiles with subtracted background is negligible compared to the noise, so that no further information is added when making the profile vector bigger. The FFT algorithm is then applied to the zero-padded profile. The resulting curve is fitted numerically, using the FFT of a circular wave that is truncated at $x_b$, to which a small offset 
accounting for the experimental noise is added. 

\begin{acknowledgement}

The authors thank E. Schatz from Nanostruc GmbH for the fabrication of the sample and for valuable discussions. The authors also thank L. Giannini for his help with the s-SNOM measurements at different incident angles. L.N.C. thanks M. Fischer and M. A. Jørgensen for valuable discussions about the fitting methods and theory. 
This work was funded by the Danish National Research Foundation through the Center for Nanostructured Graphene (grant number DNRF103) and through NanoPhoton - Center for Nanophotonics (grant number DNRF147). L.N.K. and K.J.K. acknowledge support from the Center for Nanostructured Graphene (DNRF103). N.S. acknowledges support from VILLUM FONDEN (grant no. 00028233). M.W. and N.S. acknowledge support from the Independent Research Fund Denmark - Natural Sciences (project no. 0135-00403B). S.X. acknowledges the support from Independent Research Fund Denmark (project no. 9041-00333B).

\end{acknowledgement}

\begin{suppinfo}

Additional measurements for different angles $\varphi$, derivation of the proposed model, discrete and analytical Fourier transform of a circular wave and fitting of additional measurements at grazing angle.

\end{suppinfo}

\bibliography{achemso-demo}


\newpage 
\begin{center}
    \title{\LARGE \sffamily \textbf{Supplementary information: Quantitative \\ \vspace{0mm} near-field characterization of surface plasmon \\ \vspace{5mm} polaritons on monocrystalline gold platelets}}
\end{center}
\maketitle
\renewcommand{\thefigure}{S\arabic{figure}}
\renewcommand{\thetable}{S\arabic{table}}
\renewcommand{\theequation}{S\arabic{equation}}



\maketitle

\section{S1: Angle dependency}

\begin{figure} [H]
  \centering
  \includegraphics[width=1\textwidth]{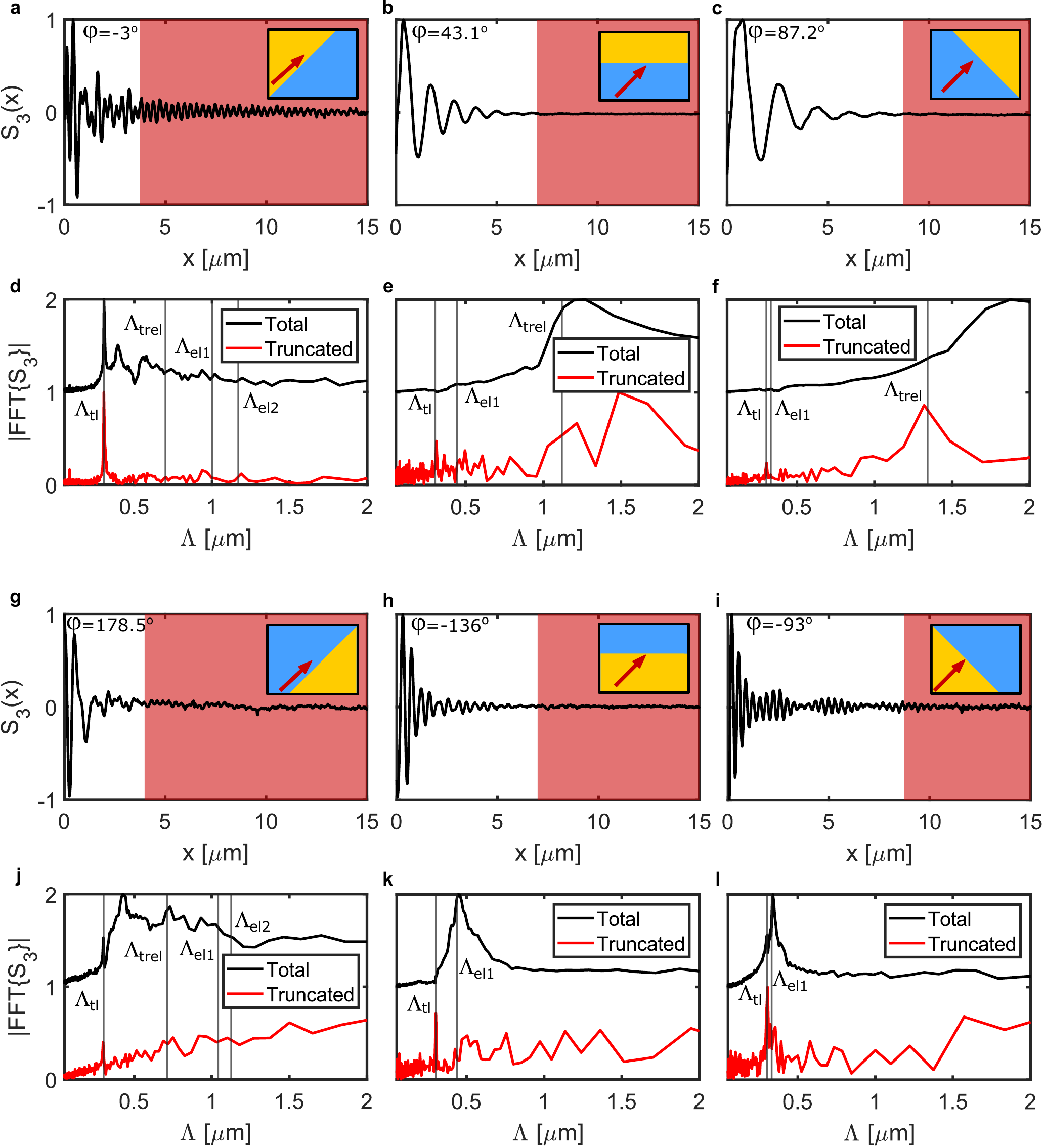}
  \caption{Angle dependency of the different SPPs. (a)-(c) and (g)-(i) show the near-field amplitude profiles of the scanned region on the gold platelets for different angles $\varphi$. The red areas indicate the portion of the curve where the tip-launched plasmons dominate. (d)-(f) and (j)-(l) represent the  Fast Fourier transforms (FFT) of (a)-(c) and (g)-(i), respectively, as a function of the wavelength $\Lambda = 2\pi/K$. The black curves in (a)-(c) and (g)-(i) are the FFT of the total data, starting at $x=0$, and the red curves the FFT of the data in the red areas. For clarity the FFT of the total data is vertically shifted.}
  \label{fig:1S} 
\end{figure}

Figure \ref{fig:1S} shows the profiles and respective Fourier transform extracted from six measurements on the same edge of the same platelet, where the azimutal angle $\varphi$ between the edge and the incident light has been varied. The different types of edge-launched, tip-reflected edge-launched and tip-launched SPPs are visible. The tip-launched SPPs are the only ones that do not have an angle-dependent wavelength. They can be seen most clearly when exciting at grazing angle. The wavelength of the tip-reflected edge-launched plasmon wavelength depends on the tip edge angle, which can slightly vary from one tip to another. Here, the wavelengths of the tip-reflected edge-launched SPPs could be recognized for a value of the polar angle $\theta'$ - defined in the main article - of $\theta'=35^\circ$.

\section{S2: Derivation of the model}

\subsection{S2.1: Measurement profiles}
The total near-field obtained with the s-SNOM measurements can be expressed as
\begin{equation}
    E_{NF} = E_{tl} + E_g ~ \text{with} ~ E_{tl} = \frac{A}{\sqrt{x}} e^{-\Gamma x}  e^{i(K_{tl}x-\phi_p)} ~ \text{and} ~ E_g = |E_g|e^{i\phi_g},
\end{equation}
with $E_{tl}$ the tip-launched SPP modelled as a circular wave, and $E_g$ a field approximated as a constant. From these expressions, it can be deduced that far from the edge, where the contribution from the tip-launched SPPs becomes negligible,  $E_{NF} = E_g$. The value of $|E_g|$ can thus be deducted from the amplitude profile.

As for the phase profile, when using the pseudo-heterodyne detection, the measured phase is the difference between the actual near-field phase and a constant reference phase $\Phi_R$ corresponding to the average optical path difference between the near-field signal and the reference \cite{Ocelic2006pseudoheterodyne,Atkin2012nano}. The measured phase, called $\Phi_{NF}$, is thus known up to the constant $\Phi_R$ and can be expressed as
\begin{equation} \label{eNA}
    \Phi_{NF} = \arctan\left( \frac{\Im(E_{tl} + E_g)}{\Re(E_{tl} + E_g)} \right) - \Phi_R.
\end{equation} 

\subsection{S2.2: Amplitude}
The absolute value squared of the total field can be expressed as 
\begin{align}
    |E_{NF}|^2 &= |E_{tl} + E_g|^2 \label{e2} \\ 
               &= |E_{tl}|^2 + |E_g|^2 + E_{tl}^*E_g + E_{tl}E_g^*\\
               &= \frac{A^2}{x}e^{-2\Gamma x} +  |E_g|^2+ 2\frac{A |E_g|}{\sqrt{x}} e^{-\Gamma x} \cos(K_{tl}x-\phi_p- \phi_g). \label{e4}
\end{align}
Checking the experimental amplitude profile from the main text and fitting with Eq. \ref{e2}, we obtain: $|E_g|\approx 5.1$ mV, $A \approx 0. 13$ mV.$\sqrt{\text{\textmu m}}$. In addition, $x$ is in the interval [3~\textmu m,~30~\textmu m]. Thus, in the fitted interval, the maximum value of the dimensionless quantity $\frac{A}{|E_g|\sqrt{x}}$ is $\frac{A}{|E_g|\sqrt{x}} \approx \frac{0.13 \ mV.\sqrt{\mu m}}{5.1 \ mV . \sqrt{3} \sqrt{\mu m}} \approx 0.015$. 

\noindent To first-order Taylor approximation in the small variable $\frac{A}{|E_g|\sqrt{x}}$, the first term of Equation \ref{e4} can be left out and this leads to
\begin{align}
    |E_{NF}|^2 &\approx |E_g|^2 \left( 1 + 2\frac{A}{|E_g|\sqrt{x}} e^{-\Gamma x} \cos(K_{tl}x-\phi_p- \phi_g) \right).
\end{align}
So the amplitude can be expressed as
\begin{equation}
    |E_{NF}| = |E_g| \sqrt{1 + 2\frac{A}{|E_g|\sqrt{x}} e^{-\Gamma x} \cos(K_{tl}x-\phi_p- \phi_g)}.
\end{equation}
Using again the first-order Taylor approximation, we find our final expression for the amplitude profile
\begin{empheq}[box=\mymath]{equation} \label{eA}
    |E_{NF}| \approx |E_g| + \frac{A}{\sqrt{x}} e^{-\Gamma x} \cos(K_{tl}x-\phi)
\end{empheq}
with $\phi = \phi_p + \phi_g$.
\bigskip 

The information to which we have access in the amplitude fit is thus $|E_g|$, $A$, $\phi$, and the SPP characterization constants $K_{tl}$ and $\Gamma$. 
As we have only access to a global phase $\phi= \phi_p + \phi_g$ in the total near-field amplitude measurement, we should express the total near-field phase in the same way in the following section, to look for the $\pi/2$ shift observed between the amplitude and phase profiles.

\subsection{S2.3: Phase}

\textbf{NB:} All the following considerations suppose $\Re(E_g)>0$, meaning that the phase profile has values always between $-\pi/2$ and $\pi/2$. Having $\Re(E_g)<0$ would only change the result by an additional constant $\pi$, so the analysis would be the same. 
\bigskip

\noindent In this section, we will derive the phase profile expression for two different cases: 
\begin{itemize}
    \item for $\Re(E_g)$ not small compared to $|E_g|$, meaning $\Re(E_g) \gg \Im(E_g)$ (so $\Re(E_g) \approx |E_g|$) or $\Re(E_g) \approx \Im(E_g)$ (so $\Re(E_g) \approx |E_g|/\sqrt{2}$)
    \item for $\Re(E_g)$ small compared to $|E_g|$, meaning $\Re(E_g) \ll \Im(E_g)$.
\end{itemize}

\subsubsection{S2.3.1: Case where $\Re(E_g)$ is not small compared to $|E_g|$}
Let us suppose that $\Re(E_g)$ is not small compared to $|E_g|$, ie $\Re(E_g) \gg \Im(E_g)$ or $\Re(E_g) \approx \Im(E_g)$. To simplify the expressions, we calculate the quantity $F = \tan(\Phi_{NF}+\Phi_R)= \frac{\Im(E_{tl} + E_g)}{\Re(E_{tl} + E_g)}$.
This quantity F can be expressed as
\begin{align}
    F &= \frac{\Im(E_g)+|E_p| \sin(K_{tl}x-\phi_p)}{\Re(E_g)+|E_p| \cos(K_{tl}x-\phi_p)} \label{e8} \\
    &= \frac{1}{\Re(E_g)} \frac{\Im(E_g)+|E_p| \sin(K_{tl}x-\phi_p)}{1+|E_p|/\Re(E_g) \cos(K_{tl}x-\phi_p)} \label{e9}
\end{align}
with $|E_p|=\frac{A}{\sqrt{x}} e^{-\Gamma x}$.
\bigskip 

\noindent Since $|E_p|/\Re(E_g) = \frac{A}{\Re(E_g) \sqrt{x}} e^{-\Gamma x} \ll 1$, we have at the first order approximation
\begin{align}
    F &\approx \frac{1}{\Re(E_g)}[\Im(E_g)+|E_p| \sin(K_{tl}x-\phi_p)][1-|E_p|/\Re(E_g) \cos(K_{tl}x-\phi_p)]\\
      &\approx \frac{\Im(E_g)}{\Re(E_g)} + \frac{|E_p|}{\Re(E_g)}\sin(K_{tl}x-\phi_p)-\frac{\Im(E_g)}{\Re(E_g)^2} |E_p|\cos(K_{tl}x-\phi_p).
\end{align}
Using the polar representation of $E_g = |E_g|e^{i\phi_g}$, one can rearrange the terms and obtain
\begin{align}
    F &= \frac{1}{\Re(E_g)^2} \left[\Im(E_g)\Re(E_g) + |E_g||E_p|\sin(K_{tl}x-\phi_p-\phi_g)\right]\\
      &= \frac{1}{\Re(E_g)^2} \left[\Im(E_g)\Re(E_g) + |E_g|\frac{A}{\sqrt{x}}e^{-\Gamma x}\sin(K_{tl}x-\phi)\right].
\end{align}
This means that our phase $\Phi_{NF}$ can be approximated as 
\begin{equation}
    \label{eP1}
    \Phi_{NF} \approx \arctan\left(\frac{\Im(E_g)}{\Re(E_g)} + |E_g|\frac{A}{\Re(E_g)^2\sqrt{x}}e^{-\Gamma x}\sin(K_{tl}x-\phi)\right) - \Phi_R.
\end{equation}

We can then use the first-order Taylor approximation of $\arctan(a+X)$ in the variable $X$, for $X \ll 1$ and $a=\frac{\Im(E_g)}{\Re(E_g)}$,
\begin{equation}\label{eTE}
    \arctan(a+X) \approx \arctan(a) + \frac{X}{1+a^2}
\end{equation}
to find from Equation \ref{eP1} the expression
\begin{equation}
    \Phi_{NF} \approx \arctan\left(\frac{\Im(E_g)}{\Re(E_g)}\right) - \Phi_R + \frac{|E_g|}{\Re(E_g)^2\left(1+\frac{\Im(E_g)^2}{\Re(E_g)^2}\right)}\frac{A}{\sqrt{x}}e^{-\Gamma x}\sin(K_{tl}x-\phi).
\end{equation}
It follows that
\begin{equation}\label{eP2a}
    \setlength\fboxsep{0.25cm}
    \setlength\fboxrule{0.4pt}
    \boxed{\Phi_{NF} \approx \arctan\left(\frac{\Im(E_g)}{\Re(E_g)}\right) - \Phi_R + \frac{A}{|E_g|\sqrt{x}}e^{-\Gamma x}\sin(K_{tl}x-\phi).}
\end{equation}
The sinusoidal behaviour and the $\pi/2$ phase shift compared to the amplitude profile (Equation \ref{eA}) is here clearly visible. 

\subsubsection{S2.3.2: Case where $\Re(E_g)$ small compared to $|E_g|$}
Let us consider the case for which \textbf{$|\Im(E_g)| \gg \Re(E_g)$}. For this case, we can use the relationship $\arctan(\frac{1}{X}) = \pm \frac{\pi}{2}-\arctan(X)$, where the $+$ sign corresponds to $X>0$ and $-$ sign to $X<0$. We can then get back to the same starting point as in Eq. \ref{e9}, with the imaginary parts being replaced by the real parts and inversely:
\begin{equation}
    G = \frac{1}{F} = \frac{1}{\Im(E_g)} \frac{\Re(E_g)+|E_p| \cos(K_{tl}x-\phi_p)}{1+|E_p|/\Im(E_g) \sin(K_{tl}x-\phi_p)}.
\end{equation}
Thus, the derivation is similar to the first case and we obtain
\begin{align}
    G &=  \frac{1}{\Im(E_g)^2} \left[\Re(E_g)\Im(E_g) + |E_g|\frac{A}{\sqrt{x}}e^{-\Gamma x}\sin(\phi-K_{tl}x)\right].
\end{align}
This means that in this case our phase $\Phi_{NF}$ can be expressed as
\begin{equation}
    \Phi_{NF} \approx \pm \frac{\pi}{2}-\arctan\left(\frac{\Re(E_g)}{\Im(E_g)} - |E_g|\frac{A}{\Im(E_g)^2\sqrt{x}}e^{-\Gamma x}\sin(K_{tl}x-\phi)\right) - \Phi_R
\end{equation}
and using again Equation \ref{eTE}, with in this case $a = \frac{\Re(E_g)}{\Im(E_g)}$, we find
\begin{equation}
    \Phi_{NF} \approx \pm \frac{\pi}{2} - \arctan\left(\frac{\Re(E_g)}{\Im(E_g)}\right) - \Phi_R + \frac{A}{|E_g|\sqrt{x}}e^{-\Gamma x}\sin(K_{tl}x-\phi)
\end{equation}
which can be reduced to
\begin{equation}\label{eP3}
    \setlength\fboxsep{0.25cm}
    \setlength\fboxrule{0.4pt}
    \boxed{\Phi_{NF} \approx \arctan\left(\frac{\Im(E_g)}{\Re(E_g)}\right) - \Phi_R + \frac{A}{|E_g|\sqrt{x}}e^{-\Gamma x}\sin(K_{tl}x-\phi).}
\end{equation}
The sinusoidal behaviour and the $\pi/2$ phase shift compared to the amplitude profile is also clearly visible here. 
\bigskip 

In summary, in both cases, the relative $\pi/2$ shift between the amplitude and phase profiles is present and the sinusoidal behaviour is visible. Furthermore, for both cases, the final equation for the phase is 
\begin{empheq}[box=\mymath]{equation} \label{eP4}
    \Phi_{NF} \approx \phi_g - \Phi_R + \frac{A}{|E_g|\sqrt{x}}e^{-\Gamma x}\sin(K_{tl}x-\phi).
\end{empheq} 

\subsection{S2.4: Comparison between the approximated expression of the phase and the non-approximated expression}
\begin{figure} [H]
    \centering
    \includegraphics[width=\textwidth]{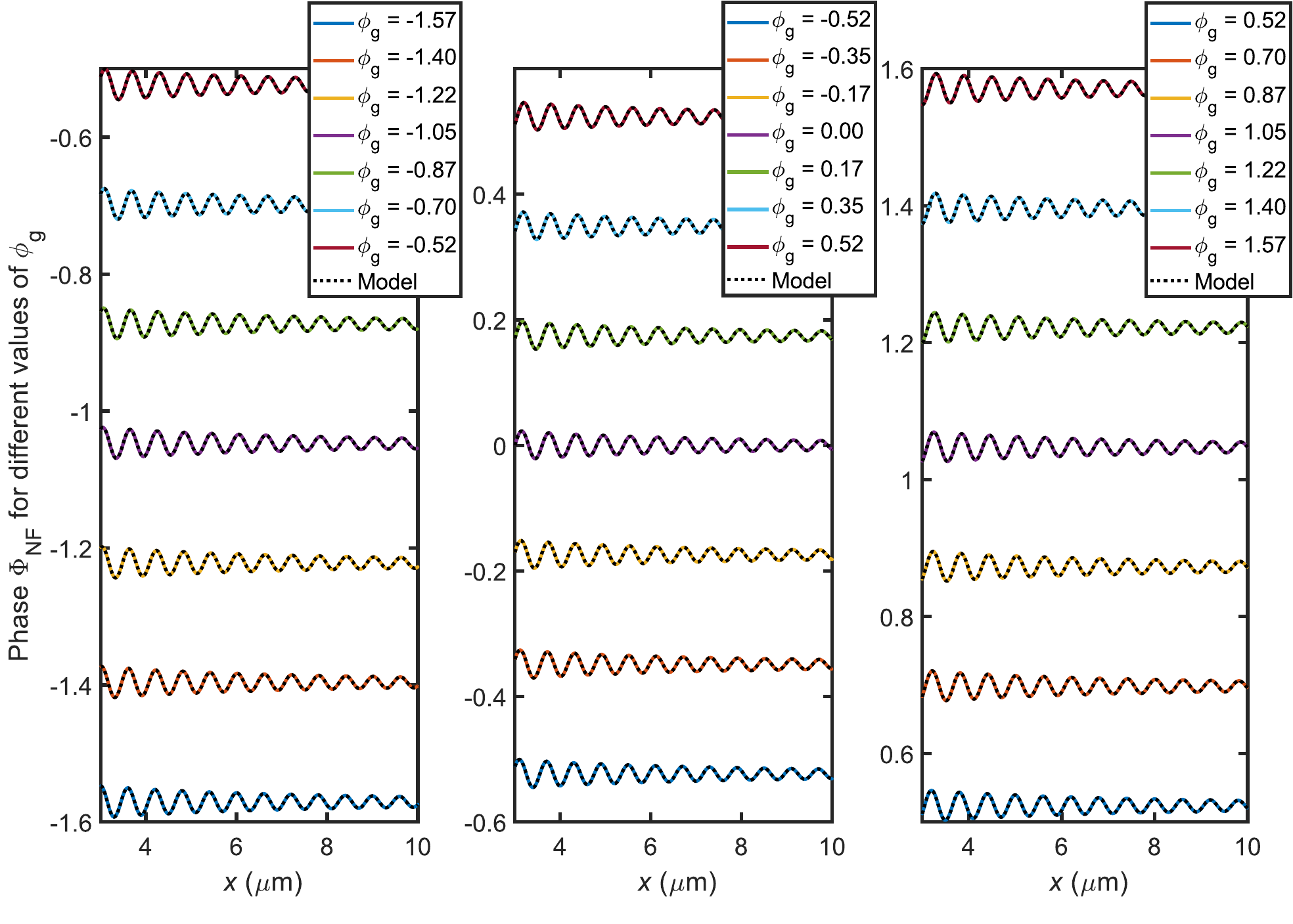}
    \caption{Comparison between the first-order approximation of the Taylor expansion of $\Phi_{NF}$ and the non-approximated expression, for values of $\phi_g$ varying from $-\pi/2$ to $\pi/2$.}
    \label{fig:CompaPhase}
\end{figure}

To verify that the first-order approximation of the phase expression is valid for all cases, we plot on the same graphs in Figure \ref{fig:CompaPhase} the first-order approximation of the Taylor expansion of the phase $\Phi_{NF}$ (Equation \ref{eP4}) - called model - and the non-approximated expression of $\Phi_{NF}$ (Equation \ref{eNA}), for different values of $\phi_g$. To simplify the expressions, we choose here $\Phi_R=0$ and $\phi_p=0$. $|E_g|$ is here taken to be 20 times larger than $A$. 

In all cases, the model and the non-approximated expressions are on top of each other.

\section{S3: Discrete and analytical Fourier transform of a circular wave}
\begin{figure}
    \centering
    \includegraphics[width=\textwidth]{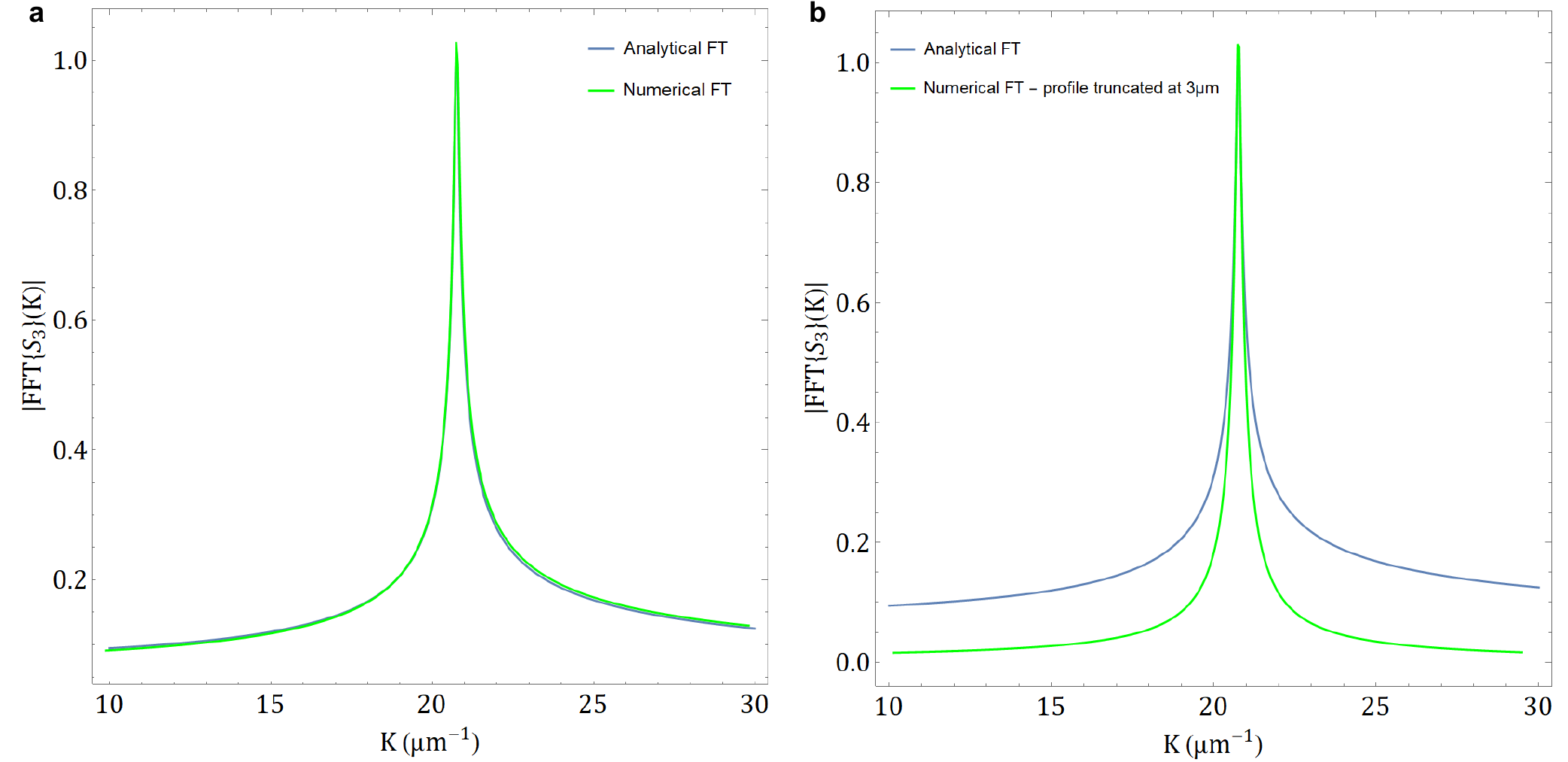}
    \caption{Comparison between the analytical and discrete Fourier transform, (a) for a circular wave profile starting at 0 \textmu m and (b) for a profile starting at 3 \textmu m.}
    \label{fig:NA}
\end{figure}

An relatively simple expression of the Fourier transform can be found using the Mathematica program. For the amplitude profile of Equation \ref{eA}, the expression is 
\begin{equation} \label{e23}
    FT\{S_3\}(K)= \frac{\sqrt{\frac{\pi}{2}}\sqrt{1+\sqrt{1-\frac{K_{tl}^2L_p^{tl^2}}{(i+L_p^{tl}K)^2}}}}{\sqrt{\frac{1}{L_p^{tl}}-iK}\sqrt{1-\frac{K_{tl}^2L_p^{tl^2}}{(i+L_p^{tl}K)^2}}}.
\end{equation}
The comparison between this expression and the corresponding discrete Fourier transform is presented in Figure \ref{fig:NA}a. The two curves are very close and thus the two fitting methods could be considered as equivalent. However, in our case, we truncate our profiles at about 3 \textmu m. The information from the profile between 0 and 3 \textmu m is thus lost. As a result, the shape of the discrete Fourier transform changes and is not well represented by Equation \ref{e23} anymore, as can be seen in Figure \ref{fig:NA}b. 

Because the analytical expression of the  Fourier transform of the truncated profile is complex and depends on the starting value $x_b$ - presented in the main text- the fit of the Fourier transform of our measurements profiles is done numerically.  

\section{S4: Fit of other measurements at grazing angle}

\subsection{Measurement from the main text: $\varphi=-3^\circ$}
To evaluate the additional uncertainty induced by the choice of $x_b$ (see main text), we have fitted this same data set for 30 different starting values of $x_b$, from $3.42$ \textmu m to $4.32$ \textmu m, and calculated the average value and variance of all these fits. The values obtained with this method are summarized in Table \ref{tab:0}. 

\begin{table}[H]
    \centering
    \resizebox{.9\textwidth}{!}{
    \begin{tabular}{l ccccc}
        \hline
        \rowcolor{Gray}
         & \multicolumn{2}{c}{\textbf{Amplitude profile}} & \multicolumn{2}{c}{\textbf{Phase profile}} & \textbf{Theory} \\
        \rowcolor{LightGray}
        & k-space & Real space  & k-space  & Real space &   \\
        $K_{\text{tl}}$ (\textmu m$^{-1}$) & $20.65 \pm 0.01$ & $20.662 \pm 0.003$ & $20.67 \pm 0.01$ & $20.67 \pm 0.01$ & 20.75 \\
        $L_\text{p}^{\text{tl}}$ (\textmu m) & $11.3 \pm 0.8$ & $12.5 \pm 0.7$ & $11.6 \pm 0.7$ & $12.1 \pm 0.7$ & 13.65 \\
        \hline
    \end{tabular}
    }
    \caption{Comparison of values for the wave-vector and propagation length, as obtained in real space and in Fourier space, for an average of 30 different starting values.}
    \label{tab:0}
\end{table}

\subsection{First additional measurement: $\varphi=-1.4^\circ$}

\begin{figure} [H]
  \centering
  \includegraphics[width=\textwidth]{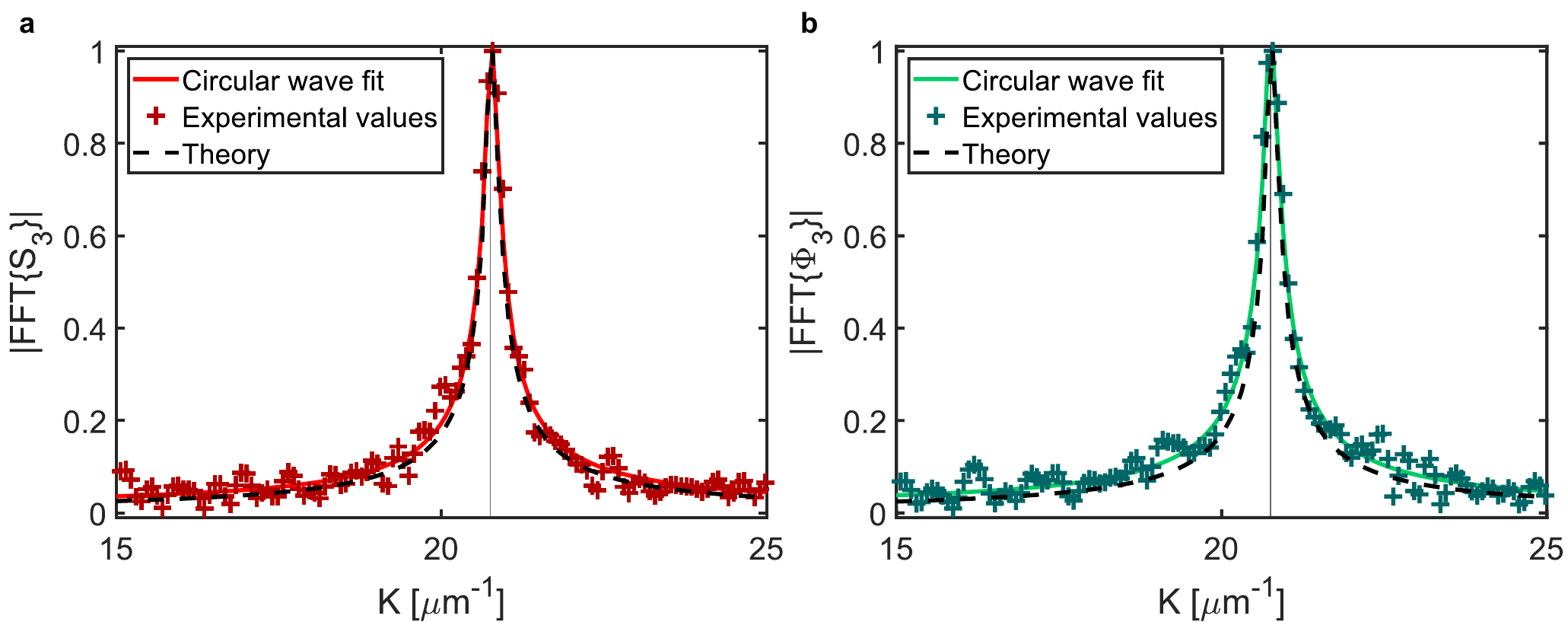}
  \caption{Fit of amplitude and phase at grazing angle ($\varphi=-1.4^\circ$), for a starting value $x_b=3$ \textmu m. (a) Fit of the tip-launched SPPs' amplitude peak. (b) Fit of the corresponding phase data.}
  \label{fig:2S}
\end{figure}

Figure \ref{fig:2S} shows the fitting in k-space of the first additional measurement, measured 8 months after the measurement shown in the main text. During these 8 months, the sample has been stored in a desiccator at room temperature.  The values obtained in Table \ref{tab:1S} still agree with those presented in the main text, with a relative change of $0.4\%$ in the mean value of $K_{tl}$ and of $0.8\%$ in the mean value of $L_p^{tl}$. This shows the durability of our monocrystalline gold sample protected by the aluminum oxide layer.  

When fitting, we find the values $|E_g| \approx 9.1$ mV and $A \approx 0.24$ mV.$\sqrt{\text{\textmu m}}$. So with $x_b = 3$~\textmu m, the maximum value of the dimensionless quantity $\frac{A}{|E_g|\sqrt{x}}$ is $\frac{A}{|E_g|\sqrt{x_b}} \approx 0.015$. The first-order Taylor expansion in this small parameter as presented in the previous section is thus still valid in this case. The values obtained for the wavevector and propagation length in the two cases presented in the main text are summarized in Tables \ref{tab:1S} (corresponding to a fixed $x_b$) and \ref{tab:2S} (averaged over $x_b$ values).

\begin{table}[H]
    \centering
    \resizebox{.9\textwidth}{!}{
    \begin{tabular}{|c||c|c||c|c||c|}
        \hline
        Profile & \multicolumn{2}{c||}{Amplitude} & \multicolumn{2}{c||}{Phase} & Theory\\
        \hline
        Fitting in: & k-space & real space  & k-space  & real space & -  \\
        \hline
        $K_{tl}$ (\textmu m$^{-1}$) & $20.752 \pm 0.007$ & $20.75 \pm 0.01$ & $20.736 \pm 0.004$ & $20.743 \pm 0.008$ & 20.75 \\
        \hline
        $L_p^{tl}$ (\textmu m) & $11.8 \pm 1$ & $14 \pm 2$ & $10.6 \pm 1$ & $11.4 \pm 0.1$ & 13.65 \\
        \hline
    \end{tabular}
    }
    \caption{Comparison of values for the wave-vector and propagation length, as obtained in real space and in Fourier space, for one starting value $x_b=3$ \textmu m.}
    \label{tab:1S}
\end{table}

\begin{table}[H]
    \centering
    \resizebox{.9\textwidth}{!}{
    \begin{tabular}{|c||c|c||c|c||c|}
        \hline
        Profile & \multicolumn{2}{c||}{Amplitude} & \multicolumn{2}{c||}{Phase} & Theory \\
        \hline
        Fitting in: & k-space & real space  & k-space  & real space & -  \\
        \hline
        $K_{tl}$ (\textmu m$^{-1}$) & $20.752 \pm 0.006$ & $20.754 \pm 0.004$ & $20.740 \pm 0.008$ & $20.745 \pm 0.004$ & 20.75 \\
        \hline
        $L_p^{tl}$ (\textmu m) & $11.7 \pm 0.3$ & $13.6 \pm 0.8$ & $10.7 \pm 0.2$ & $11.3 \pm 0.4$ & 13.65 \\
        \hline
    \end{tabular}
    }
    \caption{Comparison of values for the wave-vector and propagation length, as obtained in real space and in Fourier space, for an average of 30 different starting values.}
    \label{tab:2S}
\end{table}

\newpage
\subsection{Second additional measurement: $\varphi=-0.4^\circ$}
The second additional measurement has also been made 8 months after the measurement shown in the main text. 

\begin{figure} [H]
  \centering
  \includegraphics[width=\textwidth]{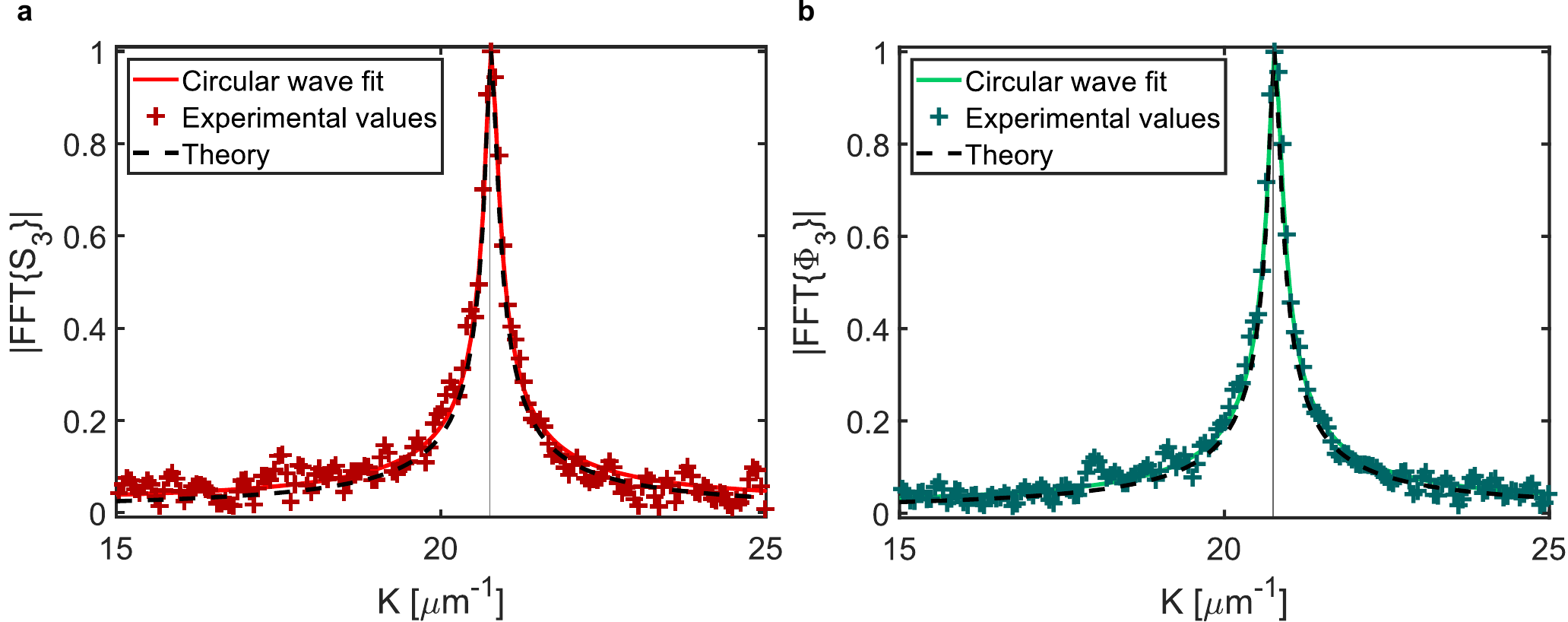}
  \caption{Fit of amplitude and phase at grazing angle ($\varphi=-0.4^\circ$), for a starting value $x_b=3$ \textmu m. (a) Fit of the tip-launched SPPs' amplitude peak. (b) Fit of the corresponding phase data.}
  \label{fig:3S}
\end{figure}

Here, in the second additional measurement, with $\varphi = -0.4$, we find the values $|E_g| \approx 23.4$ mV and $A \approx 0.47$ mV.$\sqrt{\text{\textmu m}}$. So with $x_b = 3$  \textmu m, the maximum value of the dimensionless quantity $\frac{A}{|E_g|\sqrt{x}}$ is around $0.012$, meaning that the first-order Taylor approximation is again still valid. The results are summarized in Tables \ref{tab:3S} (corresponding to a fixed $x_b$) and \ref{tab:4S} (averaged over $x_b$ values). The values obtained in Table \ref{tab:3S} also agree with those presented in the main text, with a relative change of $0.4\%$ in the mean value of $K_{tl}$ and of $8\%$ in the mean value of $L_p^{tl}$.

\begin{table}[H]
    \centering
    \resizebox{.9\textwidth}{!}{
    \begin{tabular}{|c||c|c||c|c||c|}
        \hline
        Profile & \multicolumn{2}{c||}{Amplitude} & \multicolumn{2}{c||}{Phase} & Theory\\
        \hline
        Fitting in: & k-space & real space  & k-space  & real space & -  \\
        \hline
        $K_{tl}$ (\textmu m$^{-1}$) & $20.757 \pm 0.005$ & $20.76 \pm 0.01$ & $20.754 \pm 0.008$ & $20.76 \pm 0.01$ & 20.75 \\
        \hline
        $L_p^{tl}$ (\textmu m) & $12.0 \pm 0.9$ & $13.2 \pm 0.3$ & $12.7 \pm 1$ & $14.4 \pm 3$ & 13.65 \\
        \hline
    \end{tabular}
    }
    \caption{Comparison of values for the wave-vector and propagation length, as obtained in real space and in Fourier space, for one starting value $x_b=3$ \textmu m.}
    \label{tab:3S}
\end{table}

\begin{table}[H]
    \centering
    \resizebox{.9\textwidth}{!}{
    \begin{tabular}{|c||c|c||c|c||c|}
        \hline
        Profile & \multicolumn{2}{c||}{Amplitude} & \multicolumn{2}{c||}{Phase} & Theory \\
        \hline
        Fitting in: & k-space & real space  & k-space  & real space & -  \\
        \hline
        $K_{tl}$ (\textmu m$^{-1}$) & $20.758 \pm 0.007$ & $20.759 \pm 0.007$ & $20.758 \pm 0.007$ & $20.758 \pm 0.003$ & 20.75 \\
        \hline
        $L_p^{tl}$ (\textmu m) & $12.4 \pm 0.5$ & $14.8 \pm 0.9$ & $12.0 \pm 0.6$ & $13.0 \pm 0.4$ & 13.65 \\
        \hline
    \end{tabular}
    }
    \caption{Comparison of values for the wave-vector and propagation length, as obtained in real space and in Fourier space, for an average of 30 different starting values.}
    \label{tab:4S}
\end{table}

In summary, the additional measurements are in agreement with the results presented in the main text. Averaging over different starting values gives values close to the values found for a fixed starting value $x_b$. The fit is thus stable with regards to the choice of $x_b$. Thus, averaging over different starting points, while providing relevant information about the stability of the fit, is not a necessity. 



\end{document}